\documentclass[12pt]{article}
\usepackage{times}
\usepackage[a4paper,body={16cm,23cm},left=2.5cm,right=2.5cm,top=4cm]{geometry}
\usepackage{amsmath}
\usepackage{graphicx}
\usepackage{subfigure}
\usepackage{psfrag}
\usepackage{subfigure}
\usepackage{color}
\usepackage{verbatim}
\usepackage{float}
\usepackage{url}
\usepackage[latin1]{inputenc}
\usepackage[T1]{fontenc} 
\usepackage[french]{babel}

\newcommand{\be}{\begin{equation}}
\newcommand{\ee}{\end{equation}}
\newcommand{\bi}{\begin{itemize}}
\newcommand{\ei}{\end{itemize}}
\newcommand{\bc}{\begin{center}}
\newcommand{\ec}{\end{center}}

\newcommand{\bn}{\begin{enumerate}}
\newcommand{\en}{\end{enumerate}}
\def\t2#1{\underline{\underline{#1}}}

\def\th\lim{\displaystyle \lim_{th}}

\newtheorem{rmq}{Remarque}
  {\end{rmq}%
  \end{center}}
\renewcommand{\author}[1]{
\begin{center}
{\bf #1}
\end{center}
\par
}
\renewcommand{\title}[1]{\begin{center}
    {\large{\bf #1}}
     \end{center}
\medskip
} 

\catcode`\@=11
\def\import{\begingroup
  \protected@edef\@tempa{\endgroup
    \noexpand\@import{\@ifundefined{input@path}{}{\input@path}}%
    {\@ifundefined{Ginput@path}{}{\Ginput@path}}}%
  \@tempa}
\def\@import#1#2#3#4{%
  \def\input@path{{#3}#1}\def\Ginput@path{{#3}#2}%
  \input{#4}%
  \def\input@path{#1}\ifx\input@path\@empty \let\input@path\@undefined \fi
  \def\Ginput@path{#2}\ifx\Ginput@path\@empty \let\Ginput@path\@undefined \fi
}
\catcode`\@=12

\begin{document}
\pagestyle{plain}
\title{ÉTATS DE COMPACITÉ MAXIMALE POUR LES MÉLANGES BINAIRES
DE GRAINS SPHÉRIQUES : ÉTUDE PAR SIMULATION NUMÉRIQUE}
\author{Jean-Noël Roux$^{(a)}$, François Chevoir$^{(a)}$ et Fabrice Toussaint$^{(b)}$}
\bc
(a) Laboratoire des Matériaux et des Structures du Génie Civil\footnote{Laboratoire mixte, rattaché au
LCPC, à l'ENPC et au CNRS.}, Institut Navier\\
2 Allée Kepler, 77420 Champs-sur-Marne
\ec
\bc
(b) Lafarge, Centre de Recherches \\
95 rue du Montmurier BP 15\\
38291 Saint-Quentin Fallavier\\
\ec
\bc
\begin{minipage}{16cm}
{\small
   \parindent=0pt
    {\bf R\'ESUM\'E : }
Nous appliquons la simulation numérique (par la méthode des éléments discrets, également appelée dynamique moléculaire) à l'étude des assemblages
désordonnés de compacité maximale de particules sphériques, monodispersées ou bien en mélanges binaires, le rapport des diamètres étant fixé à trois.
La simulation conduit à une définition pratique de la compacité maximale des systèmes désordonnés, 
comme celle des configurations mécaniques d'équilibre sous chargement
isotrope de particules rigides non frottantes. Une étude statistique des effets de taille d'échantillons permet d'évaluer, 
dans la limite des grands systèmes, les compacités
maximales des assemblages monodispersés, pour lesquels on retrouve la valeur classique $0.639$, 
et bidispersés, pour deux valeurs distinctes de la fraction volumique de grosses
particules. Les calculs mettent en évidence l'influence sur l'état final et sa compacité de la durée de 
l'étape initiale de l'assemblage, dans laquelle les grains sont agités
et interagissent par collisions. L'allongement de cette phase d'agitation entraîne une évolution graduelle de 
l'état du système dans laquelle l'augmentation de la compacité
s'accompagne de la nucléation d'un ordre cristallin dans le cas des billes de même taille et d'une ségrégation 
progressive selon la taille dans le cas des mélanges binaires.
Même si cette évolution est lente et son effet probablement peu sensible dans nombre de situations pratiques, 
il convient de définir la compacité maximale des états désordonnés comme
celle que l'on obtient dans la limite des assemblages rapides. 
Nous indiquons pour finir quelques prolongements possibles de cette étude préliminaire de la compacité des mélanges.  
\par
}
\end{minipage}
\ec

\section{Introduction}
De nombreux matériaux du génie civil sont composés d'un assemblage de grains solides, seuls (sols pulvérulents, graves non traitées, granulats, ballasts), en présence d'un
fluide (sols saturés, ciments frais, matériaux de chaussées lors de leur mise en place), ou inclus dans une matrice solide (bétons, enrobés bitumineux en service). 
La compacité -- c'est-à-dire la fraction $\Phi$ du volume de matériau occupée par les particules solides -- 
est toujours au nombre des grandeurs utilisées pour caractériser l'état de ces matériaux, pour tenter
de prédire leur comportement ou aider à leur formulation. Souvent on se réfère à une \emph{compacité maximale}
$\Phi^*$, implicitement supposée liée à la seule géométrie des
grains. Ainsi, pour la formulation des ciments~\cite{FdL99}, on corrèle la viscosité d'une suspension à la compacité maximale des grains solides. En mécanique des sols, 
où on utilise souvent l'indice des vides $e = \frac{1-\Phi}{\Phi}$ au lieu de la compacité, des procédures normalisées en donnent des valeurs maximales et minimales, tandis
que l'étude du comportement mécanique des matériaux granulaires souligne l'intérêt de la notion de compacité (ou d'indice des vides) \emph{critique}, qui est approchée
dans les états d'écoulement plastique en grandes déformations~\cite{DMWood}.

Les méthodes de simulation discrète~\cite{RoCh05}, qui connaissent actuellement un grand essor, donnent accès à l'ensemble des degrés de liberté d'un assemblage de grains
solides. Le présent article a pour but de montrer qu'elles permettent de préciser la notion d'état de compacité maximale, 
en dégageant les influences de la géométrie et de la micromécanique,
et de déterminer $\Phi^*$ pour des systèmes simples dont la granulométrie n'est pas trop étalée. 

Pour un rapport des diamètres établi à trois, on a choisi d'étudier
des mélanges avec deux proportions différentes $p$ de volume solide
correspondant aux grosses particules~: $p=0.5$ et $p=0.7$, et de
limiter le nombre total $N$ de particules à quelques milliers dans les
échantillons simulés.

Le bilan de cette étude préliminaire, avec les
résultats obtenus, l'influence de $N$ et les
fluctuations statistiques entre échantillons différents, est présenté
et brièvement discuté dans le §\ref{sec:res1}. Auparavant, le §\ref{sec:defmeth}
donne quelques définitions utiles et présente rapidement les méthodes numériques employées, tandis que des résultats connus dans
le cas du système monodispersé sont rappelées au §\ref{sec:mono}. Nous terminons en suggérant
(§\ref{sec:persp}) quelques perspectives quant à la poursuite de
l'application de la simulation numérique discrète aux assemblages
granulaires polydispersés.

\section{Définitions, méthodes numériques.\label{sec:defmeth}}
\subsection{Qu'est-ce qu'un état de compacité maximale ?}
Avec des billes de la même taille, la compacité maximale de l'assemblage aléatoire
est classiquement estimée à une valeur légèrement inférieure à 0.64~\cite{CC87,BH93}. 

La notion d'état désordonné de compacité maximale est \emph{a priori} un peu floue. Il
s'agit en principe d'un maximum géométrique, concernant des particules
rigides et impénétrables, mais qui exclut certaines
configurations considérées comme trop ordonnées (en particulier celles
dans lesquelles on peut clairement déceler un arrangement cristallin,
au moins localement).

Une remarque permet de préciser cette définition~: si l'on se donne
une définition de la compacité $\Phi$ adaptée aux conditions aux limites
choisies, les configurations qui réalisent un maximum local de $\Phi$,
sous les contraintes d'impénétrabilité,
coïncident exactement avec les configurations d'équilibre stable de grains
rigides non frottants soumis à une pression isotrope~\cite{JNR2000}. <<Maximum local~>>
signifie qu'il n'existe pas de mouvement augmentant $\Phi$ qui ne
viole pas les contraintes stériques. Le problème posé en termes
géométriques a donc naturellement un sens mécanique~: partout où les
contraintes d'impénétrabilité sont actives, apparaît une force normale entre les
deux grains en contact, qui est un multiplicateur de Lagrange associé à
cette contrainte. 

Il est donc tout naturel de chercher à obtenir ces
états en soumettant des grains non frottants à une pression
isotrope. C'est un procédé fréquemment employé en simulation numérique
pour obtenir des configurations denses~\cite{TH00,MJS00,AgRo05,iviso}.

{\em Un état de compacité maximale est un état d'équilibre d'en ensemble de grains rigides sans
frottement dans les contacts, soumis à une pression isotrope.}
\subsection{Qu'est-ce qu'un état \emph{désordonné} de compacité maximale ?}
On dispose de différents indicateurs (paramètres d'ordre) pour la
cristallisation des systèmes de << sphères dures~>>~\cite{VCKB02,LGROT02,iviso}. Celle-ci se
manifeste à partir d'une compacité d'environ $0.49$ en présence d'une
agitation thermique maintenue (dans le modèle de fluide pour lequel
les collisions conservent l'énergie cinétique), mais elle est
relativement facile à éviter dans la pratique des calculs si on se
place à $\Phi \ge 0.55$~\cite{VCKB02}. Avec des billes macroscopiques qui dissipent
l'énergie dans leurs contacts et échappent à l'agitation thermique spontanée, on peut retrouver la tendance à la
cristallisation lorsque l'assemblage est soumis à des vibrations ou à un cisaillement oscillant~\cite{NDP00}.  

Le phénomène de cristallisation disparaît lorsque la
largeur de la distribution des diamètres (écart-type d'une
distribution gaussienne rapporté à la valeur moyenne) est de l'ordre de 6\% du
diamètre moyen~\cite{PeterBolhuis}.

Ce qu'il faut craindre alors, c'est la ségrégation des
particules par taille, autre cause possible d'inhomogénéité. Les
mélanges de sphères de différentes tailles, si on les considère comme des molécules d'un fluide 
ou des particules colloïdales soumises au mouvement brownien, peuvent exhiber un diagramme de phase
assez complexe, qui dépend fortement du rapport des tailles, comporte éventuellement plusieurs phases
fluides et plusieurs phases solides, etc...
On pourra par exemple consulter la réf.~\cite{DiRoEv99}, et la copieuse littérature à laquelle
renvoie cette étude. \`A densité élevée, la structure la plus stable implique toujours 
une séparation de deux phases de compositions différentes. Comme dans le cas monodispersé, 
en revenant aux particules macroscopiques (c'est-à-dire sans agitation thermique ou mouvement brownien),
on observe que la manipulation d'un mélange de particules de tailles différentes (qu'il s'agisse
d'une vibration continue, de secousses répétées séparées par des temps de repos, etc...) 
tend à induire le même type d'organisation que l'agitation thermique spontanée
des systèmes moléculaires ou colloïdaux~: on assiste à une ségrégation progressive~\cite{ReMu02,AuScKrRe03,KU04}. De même la mise en écoulement des matériaux granulaires
conduit à la séparation des populations de grains selon leurs tailles~\cite{RRNC06}. 

Avec les sphères d'une seule taille on peut définir un état de compacité
maximale aléatoire si on a un maximum local de $\Phi$ sous les contraintes d'impénétrabilité dans l'espace des configurations et si on ne
décèle aucun germe de cristallisation locale. 

Il est logique d'adopter une définition
analogue pour les systèmes bidispersés, en remplaçant la
cristallisation par la ségrégation.

On constate que, pour des échantillons assez grands, les états désordonnés de compacité maximale possèdent tous (\emph{approximativement},
et non pas \emph{exactement}, en raison des évolutions lentes vers la cristallisation ou la ségrégation) 
la même compacité $\Phi^*$, qui est celle que nous étudions
ici. 

Les résultats rappelés ci-dessous dans le cas des billes d'une seule taille et présentés au §\ref{sec:res1} 
pour les mélanges binaires indiqueront dans quelle mesure,
pour le procédé d'assemblage indiqué au §\ref{sec:proc} ci-dessous, la compacité peut dépendre de la durée de la phase initiale dans laquelle les grains sont agités.

{\em Un état désordonné de compacité maximale peut être défini comme un état d'équilibre d'en ensemble de grains rigides sans
frottement dans les contacts, soumis à une pression isotrope, 
pour lequel les évolutions vers l'ordre cristallin ou vers la ségrégation des mélanges restent négligeables.}

\subsection{Procédure de simulation\label{sec:proc}}
Pour produire des états homogènes, commodes à caractériser et
dépourvus d'effets de bord, on utilise des conditions aux limites
périodiques. 

La méthode de calcul utilisée est la dynamique moléculaire, appellation abrégée en DM ci-dessous, (que l'on désigne souvent comme 
<< méthode aux éléments discrets>> ) avec des sphères
assez rigides et des ingrédients d'inertie et de dissipation visqueuse
appropriés -- ces derniers garantissant une restitution nulle dans les collisions binaires. La raideur des contacts est telle que dans l'état
final l'<<interpénétration~>> typique des sphères en contact (c'est-à-dire, bien sûr, la déflexion normale du contact) soit
de l'ordre de quelques $10^{-5}d$, $d$ étant leur diamètre. On trouvera des explications et des références plus détaillées sur
la méthode dans un article récent~\cite{RoCh05}.

Au préalable, les particules sont disposées soit << au hasard >> (un
procédé qui impose de commencer les calculs dans une configuration de
compacité assez faible), soit sur les sites d'un réseau régulier.
Dans une seconde étape, la configuration de départ est mélangée~: on
donne aux grains une certaine énergie cinétique et on les laisse
interagir par des collisions qui conservent cette énergie (chocs élastiques), à la
manière des molécules d'un fluide (qui seraient toutefois dépourvues d'interaction à
distance). Lorsque la position de départ était sur un réseau, on
s'assure alors que cet ordre initial a bien disparu dans l'étape de
mélange.

La durée de la phase de mélange et son influence possible se mesurent
au nombre de collisions moyen par grain, à une compacité donnée.

Une possibilité intéressante, en particulier dans le cas de systèmes
polydispersés, pour lesquels il peut être difficile de déterminer une
configuration initiale au hasard avec une compacité qui ne soit pas
trop faible, est l'utilisation d'un
algorithme qui permette simultanément de mélanger et de compacter,
pour que le recours à la DM puisse n'intervenir que
dans une étape plus tardive. C'est ce que fait l'algorithme de
Lubachevsky et Stillinger~\cite{LS90,LSP91} (abrégé en LS ci-dessous),
qui consiste à calculer des collisions entre des objets qui se
dilatent à vitesse constante, de sorte que la compacité augmente au cours du temps.

Les résultats présentés au §\ref{sec:res1} font tous appel à cette
méthode dans une première étape, et ses effets sont testés ci-dessous
dans le cas monodispersé.
\section{Résultats pour les billes monodispersées\label{sec:mono}}
\subsection{Procédure directe, temps caractéristiques}
On a pu vérifier que l'on obtenait avec un seul diamètre de billes la
valeur 
$\Phi^*=0.637\pm 0.001$ en effectuant l'étape de mélange à $\Phi=0.45$,
(il suffit alors d'une dizaine de collisions par grain pour faire
disparaître toute mémoire de la configuration initiale) puis en
achevant le compactage par DM avec des particules
déformables et des chocs mous. Cette valeur résulte de moyennes sur une série de 5 échantillons
de N=4000 sphères. Une autre série de configurations de 1372 sphères donne 
$0.636 \pm 0.001$ \footnote{On convient ici, et dans toute la suite,
  d'indiquer comme intervalle d'incertitude un écart-type de part
et d'autre de la moyenne}. L'erreur due à la légère déformabilité des
contacts était avec le modèle retenu estimée à $\Delta \Phi=1.2 \times
10^{-4}$ (donc négligeable, vu l'incertitude sur $\Phi$).

Pour donner une idée approximative de la durée du processus de compaction (deuxième étape), qui
est calculé à pression contrôlée $P$, on peut utiliser le temps
caractéristique $\displaystyle {\tau = \sqrt{\frac{m}{dP}}}$ associé à
l'inertie des grains de masse $m$ et de diamètre $d$ soumis à des
forces d'ordre $Pd^2$. La compacité finale est approchée (à moins de
0.01 près) en quelques
dizaines de $\tau$. Ensuite, si on désire une configuration bien
équilibrée (dans laquelle le réseau des contacts et les valeurs des
forces sont déterminées avec une bonne précision), il faut poursuivre
les calculs pendant un temps qui peut atteindre quelques centaines de
$\tau$ (il faut noter qu'il est relativement difficile de stabiliser
des grains non frottants dans une position d'équilibre). 
Pour avoir un point de comparaison avec une expérience de
laboratoire, on peut considérer par exemple un processus de dépôt
sous gravité. Une telle procédure est couramment employée dans les laboratoires de mécanique des sols, et
en contrôlant débit et hauteur de chute libre -- on parle alors de \emph{pluviation contrôlée}~\cite{BENA01,RT87} --
elle constitue un procédé de fabrication reproductible d'échantillons homogènes dont la compacité peut être variée à la demande.
Par la simulation numérique~\cite{ERCCD05,ECCDR06}, on peut comparer les assemblages de billes assemblées par pluviation aux
configurations obtenues par les processus de compactage isotrope idéaux, dépourvus d'effets de bord, tels que celui que nous employons dans la présente étude.
On vérifie alors que les états que nous étudions ici sont bien similaires à ceux que l'on peut trouver dans des situations plus proches d'une 
expérience réelle.

Dans le procédé de dépôt gravitaire, la pression typique, celle qui règne dans la
couche proche de la surface libre où les grains se réarrangent,
correspond au poids d'une dizaine de couches granulaires. 
En prenant donc $P\sim 10 {\displaystyle \frac{mg}{d^2}}$, on aura
 ${\displaystyle \tau \simeq  \sqrt{\frac{d}{10g}}}$, ce qui donne environ
 $3\times 10^{-3}$s
 avec des grains millimétriques. Les temps trouvés (de l'ordre du dizième
 de seconde pour l'assemblage par déposition d'une couche de 10 grains
 d'épaisseur, ou de la seconde pour remplir un récipient de 10cm de
 hauteur) semblent donc comparables entre le calcul
 numérique et la fabrication d'échantillons en laboratoire. 

Cette compacité
pour les assemblages de sphères d'une seule
taille est en bon accord avec les résultats d'expériences en
laboratoire~\cite{CC87,BH93}, et avec d'autres travaux numériques. Ainsi
O'Hern \emph{et al.}~\cite{OSLN03} par exemple, donnent $\Phi^*=0.639\pm
0.001$, résultat obtenu par extrapolation vers la limite $N\to \infty$.
Ces auteurs ont mené à bien
une étude systématique de l'influence de la taille du système sur
la compacité maximale. Ils ont observé une légère influence du
nombre $N$ de sphères sur la valeur moyenne dans les échantillons de
$N$ sphères, $\langle \Phi \rangle
(N)$ Celle-ci tend à augmenter vers sa valeur asymptotique $\Phi^*$ dans
le limite $N\to \infty$ comme~:
\be
 \langle \Phi \rangle (N) = \Phi^* - \delta_0 N^{-\alpha},
\label{eqn:dphi}
\ee
avec $\delta _0 = 0.12\pm 0.03$ et $\alpha = 0.47 \pm 0.05$.
Ils ont également déterminé l'influence de $N$ sur la largeur $w$ (largeur
à mi-hauteur d'une courbe en cloche) de la distribution des compacités
dans les échantillons de taille $N$~:
\be
w =w_0 N^{-\Omega}, \label{eqn:wphi}
\ee
avec $w_0 = 0.16\pm 0.04$ et $\Omega=0.55 \pm 0.03$.
Les équations~\ref{eqn:dphi} et~\ref{eqn:wphi} s'appliquent aux
résultats de~\cite{OSLN03} pour $N\ge 64$. 

\subsection{Effet d'une agitation prolongée: algorithme LS}
Dans un second temps nous avons testé l'usage de
l'algorithme LS pour ne recourir à la DM qu'à une
densité plus élevée.

Si l'on utilise l'algorithme LS jusqu'à obtenir $\Phi=0.632$ (avec 6000
collisions par particule en moyenne pour arriver à cette compacité) on
constate que l'achèvement ultérieur du compactage par DM
 produit des compacités plus élevées que la valeur trouvée
précédemment~: $\Phi =0.6423 \pm 0.0004$. Il est difficile de
repérer un ordre local dans ces nouvelles configurations,
l'évolution vers l'arrangement cristallin restant très embryonnaire 
(ceci peut se contrôler quantitativement avec des
<< paramètres d'ordre >> comme dans~\cite{VCKB02,LGROT02}). Mais si on
prolonge encore l'étape de calcul par la méthode
LS (comme cela a été fait dans~\cite{TTD00,KTS02}), on peut obtenir toutes les compacités
intermédiaires entre celles que nous avons citées et la valeur
$\frac{\pi}{3\sqrt{2}}\simeq 0.74 $ du réseau cristallin parfait,
correspondant à des systèmes partiellement ordonnés selon les motifs
cubique à faces centrées ou hexagonal compact.

On voit donc qu'il
n'est pas possible en principe de considérer la valeur de la compacité
maximale $\Phi^*$ des systèmes désordonnés comme unique.
Toutefois, si l'usage de l'algorithme LS est poussé jusqu'à 
$\Phi=0.59$ avec seulement 100 collisions par bille, on obtient à
nouveau la valeur précédente ($\Phi\simeq 0.637$ pour $N=4000$,
légèrement inférieure pour $N=1372$). 

{\em On retiendra donc
que la compacité maximale la plus usuelle correspond à un
assemblage rapide, et qu'on la retrouve en limitant la durée
des phases d'agitation. }

Cette interprétation est confortée par
l'absence ou la brièveté d'une telle phase de vibration dans les procédés
pratiques d'assemblages tels que le dépôt sous gravité (pluviation) ou le damage par couches, ainsi que
par les résultats déjà cités de~\cite{OSLN03}, qui sont en très bon
accord avec les valeurs de $\Phi^*$ que nous avons obtenues en limitant la durée de l'agitation. En effet, la méthode
numérique employée par O'Hern \emph{et al.}~\cite{OSLN03} est censée correspondre à la limite
des compressions très rapides (l'énergie potentielle élastique, associée à la déformation des contacts
intergranulaires dans une configuration à volume imposé, est directement minimisée par un algorithme
de descente du gradient, et
la limite où le volume choisi donne des déformations nulles fournit la valeur de la compacité cherchée). 
\subsection{Rôle du frottement}
La procédure d'assemblage précédente, qui consiste à faire décroître le volume de la cellule (périodique) contenant les grains jusqu'à ce que le réseau des
forces de contacts vienne équilibrer chaun des grains et contrebalancer la pression extérieurement appliquée, mise en {\oe}uvre en présence de frottement
intergranulaire, conduit à des configurations mécaniquement équilibrées de compacité moindre~\cite{TH00,MJS00,AgRo05}. De plus, en présence de  frottement intergranulaire, 
on doit également noter que les grains peuvent être assemblés, pour les mêmes contraintes, dans des états assez différents, ce qui est bien sûr conforme à l'expérience, puisque
l'on caractérise classiquement l'état d'un échantillon de sable par sa compacité.

La définition des états de compacité maximale que nous avons donnée plus haut, comme états d'équilibre sous chargement isotrope de grains non frottants, conduit donc  
à considérer les procédés de compactage (vibration, cisaillement, lubrification) comme destinés à limiter ou à circonvenir les effets du frottement dans les contacts.

Pour des billes sphériques identiques, la compacité minimale que l'on peut atteindre dans une configuration d'équilibre dépend de la valeur du coefficient de frottement, et divers
procédés peuvent conduire à des valeurs différentes, les plus basses étant de l'ordre de 0,56 à 0,58 avec des coefficients de frottement compris entre 0,2 et 0,5.  

Signalons enfin que la simulation numérique a permis de constater que la compacité et le nombre de coordination d'un assemblage de billes frottantes
pouvaient varier indépendamment~\cite{RoCh05,iviso}.
\subsection{Effet d'un cisaillement \label{subsec:anis}}
Jusqu'ici nous avons considéré exclusivement des configurations
isotropes. Or, il est possible d'appliquer à des assemblages de
sphères sans frottement une contrainte de cisaillement, la
configuration restant en équilibre mécanique, avec des forces de
contact normales qui équilibrent les forces extérieurement appliquées.
Nous avons ainsi observé que l'on pouvait appliquer des contraintes
principales $\sigma _1 >\sigma_2=\sigma_3$ jusqu'à atteindre des
rapports $\sigma_1/\sigma_3$ de $1.2$ à $1.25$, avec un modèle de
sphères quasi-rigides. Lorsque le déviateur des contraintes est
augmenté, par paliers, à partir d'une configuration assemblée de façon
isotrope, on constate que le système se déforme, car le réseau des contacts
se réajuste pour pouvoir supporter les nouvelles contraintes. À
l'équilibre on obtient une distribution des orientations des contacts
anisotrope, avec un plus grand nombre de vecteurs normaux aux contacts
parallèles à la
direction principale majeure des contraintes. Il est intéressant de
constater que les réarrangements se produisent sans changement notable de la
compacité. Cette remarque nous permet donc de {\em généraliser la définition que nous avons adoptée pour un état de compacité maximale, 
en abandonnant la condition de contraintes isotropes} : c'est simplement un état d'équilibre de grains rigides non frottants. 

Dans le
langage de la mécanique des matériaux granulaires (ou des sols~\cite{DMWood}),
les assemblages de sphères non frottantes possèdent un angle de
frottement interne (faible, entre 5 et 6 degrés, mais non nul), mais sont dénués de dilatance. 
Pour un assemblage de grains \emph{parfaitement lubrifiés}, et donc non frottant, la compacité
$\Phi^*$ que
l'on évalue en imposant une pression isotrope semble coïncider avec
la <<~compacité critique~>>, celle d'un échantillon continuellement
cisaillé en régime quasi-statique.  Cette valeur $\Phi^*$ pourrait donc en principe
avoir une certaine pertinence pour l'étude de la rhéologie des pâtes, pour lesquelles
on s'intéresse à la compacité associée à la divergence de la viscosité, en régime newtonien. 
Une étude expérimentale récente~\cite{OFR06} par une équipe du LMSGC montre néanmoins que la
viscosité diverge pour des compacités inférieures à $\Phi^*$, et il semble que le comportement rhéologique des pâtes granulaires soit influencé par
l'état de surface des particules solides, que l'on ne peut pas considérer comme parfaitement lubrifiées par le fluide.

Par ailleurs, nous ignorons si
les assemblages de sphères \emph{polydispersées} non frottantes sont également dépourvus de dilatance.

Enfin, il convient de rappeler que la compacité critique des grains secs dépend du coefficient de frottement et est inférieure à la compacité maximale (elle est d'environ
0,595 pour les billes identiques lorsque le coefficient de frottement de contact vaut 0,3~\cite{RoCh05}).
\section{Résultats pour les mélanges binaires.\label{sec:res1}}
\begin{figure}[!htbp]
\centering
\includegraphics[height=14cm]{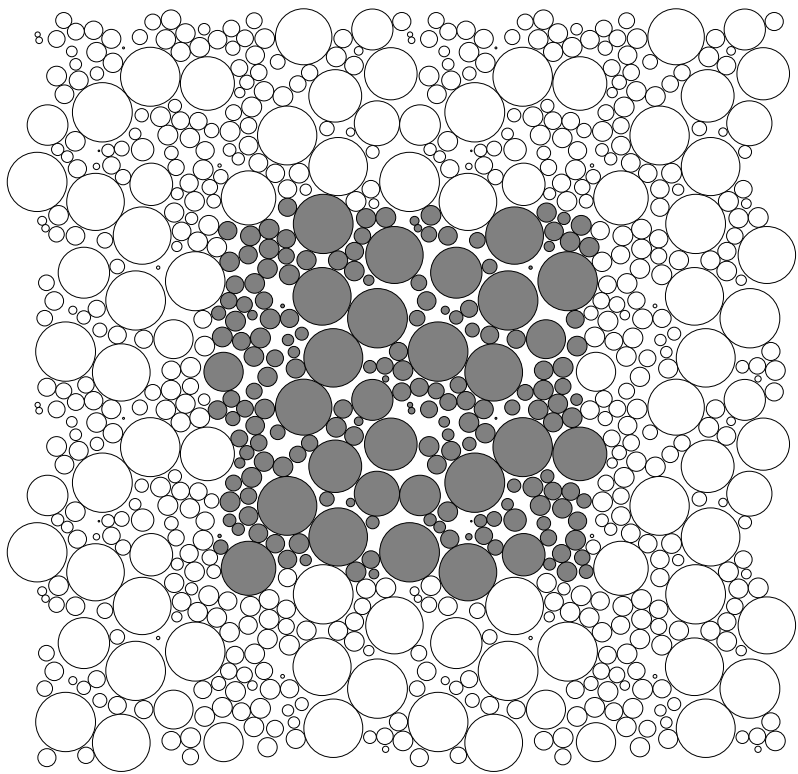}
\caption{Vue en coupe d'un échantillon avec $N=2916$ et $N_2=232$. 
Les particules figurées en gris appartiennent à la
  cellule de base dans laquelle sont conduits les calculs, et qui se
  répète périodiquement.
\label{fig:coupe}}
\end{figure}

\begin{figure}[!htbp]
\centering
\includegraphics[height=14cm]{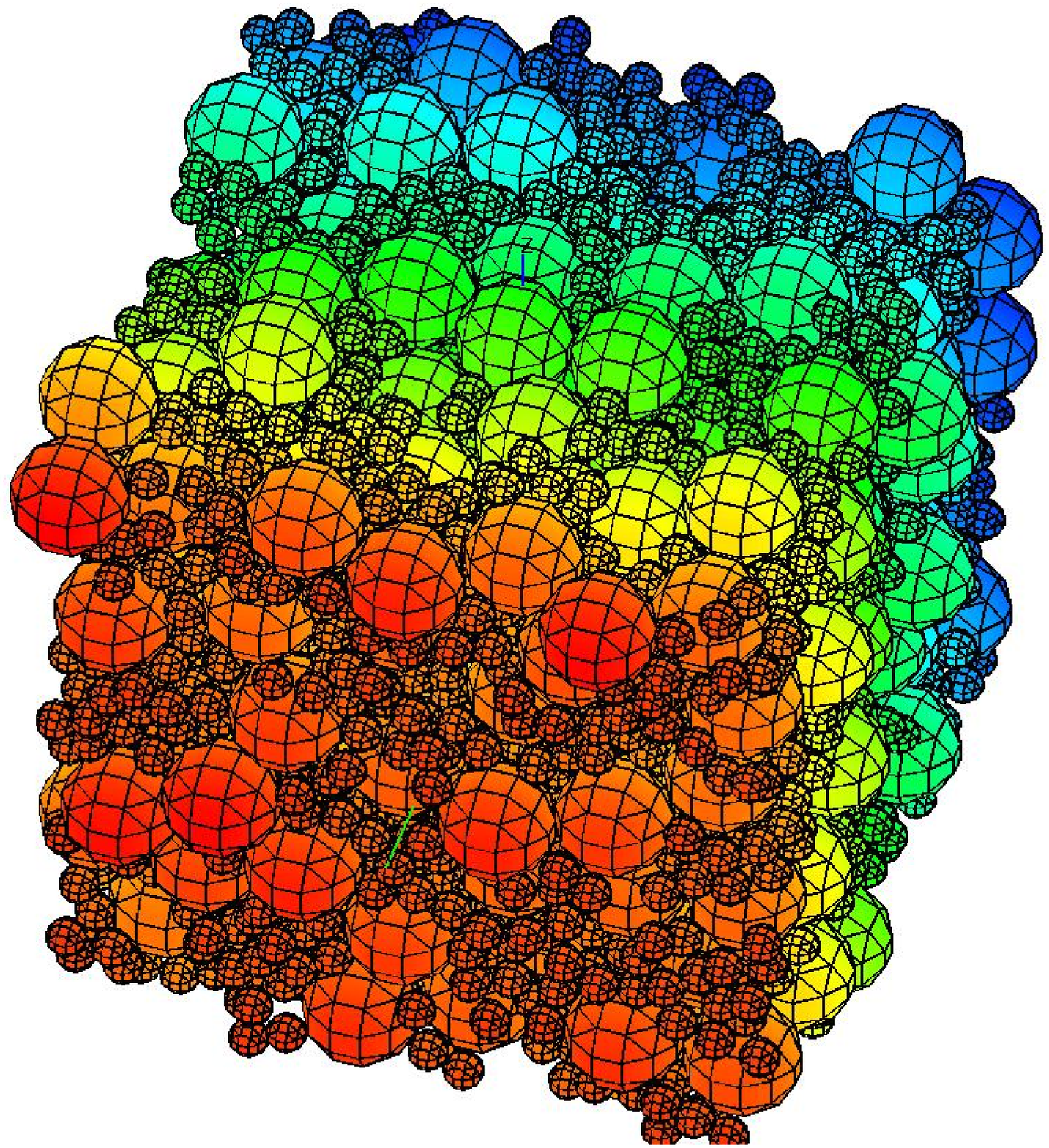}
\caption{Vue en perspective d'un échantillon avec $N=2916$ et $N_2=232$. 
\label{fig:paquet}}
\end{figure}

\begin{figure}[!htbp]
\centering
\includegraphics[height=14cm]{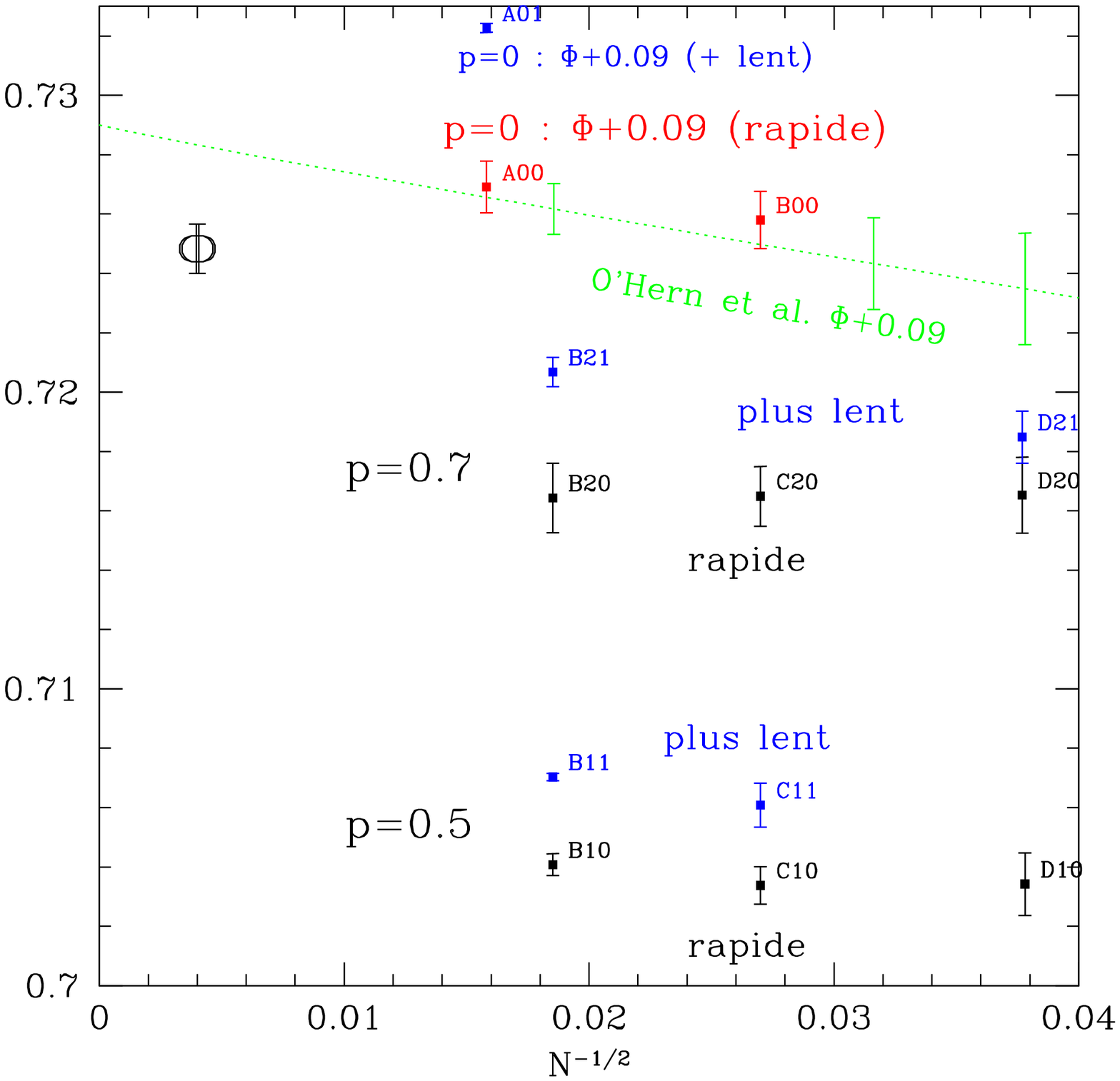}
\caption{Compacité maximale $\Phi^*$ représentée fonction de $1/\sqrt{N}$ pour les valeurs $p=0.5$ et $p=0.7$
étudiées (les barres d'erreurs s'étendent ici sur un écart-type de part et d'autre de la moyenne).  
Les codes indiqués sont ceux du tableau~\ref{tab:liste}, et on a distingué par la couleur bleue les calculs pour
lesquels une phase d'agitation prolongée a augmenté la valeur de la compacité finale, par comparaison aux
processus d'assemblages rapides. Pour faire
figurer sur le même graphe les
résultats sur les systèmes monodispersés, on a décalé les valeurs de
$\Phi^*$ correspondantes de 0.09 vers le haut. Les résultats
de~\cite{OSLN03}, résumés
par les relations~\eqref{eqn:dphi} et~\eqref{eqn:wphi}, sont figurés
en vert.
\label{fig:phi}}
\end{figure}

Nous avons simulé la compaction isotrope d'échantillons de billes de diamètre
$d$ et $D=3d$. Le tableau~\ref{tab:liste} donne la liste des différentes séries d'échantillons
ainsi préparées (il rappelle aussi les résultats obtenus au laboratoire avec les billes d'une seule
taille).  
\begin{table}[!htb]
\centering
\small
\begin{tabular}{|c||c|c|c|c|c|c|c|c|c|}  \cline{1-10}
Code & $p$ & $N$ & $N_2$ & $M$& $\Phi_0$ & $\Phi_1$ & $n_{coll}$ & $\langle \Phi^* \rangle $ & $\Delta \Phi^*$\\
\hline
\hline
A00 & 0 & 4000 & 0 &  5&  0.45 & 0.45 & 50 & $0.637 $& $ 0.001$\\
\hline
A01 & 0 & 4000 & 0 &  5&  0.05 & 0.632 & 5000 & $0.6423 $& $ 0.0004$\\
\hline
C00 & 0 & 1372 & 0 &  10&  0.45 & 0.45 & 50 & $0.636 $& $ 0.0012$\\
\hline
C01 & 0 & 1372 & 0 &  8&  0.05 & 0.59 & 80 & $0.636 $& $ 0.001$\\
\hline
\hline
B10 & 0.5 & 2916 & 104 &  5&  0.05 & 0.68 & 860 & $0.7041$ & $3\times 10^{-4}$\\
\hline
B11 & 0.5 & 2916 & 104 &  5&  0.05 & 0.70 & 4400 & $0.7070$ & $1.1\times 10^{-4}$\\
\hline
C10 & 0.5 & 1372 & 49 &  10&  0.05 & 0.68 & 850 & $0.7034$ & $ 6.1\times 10^{-4}$\\
\hline
C11 & 0.5 & 1372 & 49 &  10&  0.05 & 0.67 & 2200 & $0.7061$ & $7\times 10^{-4}$\\
\hline
D10 & 0.5 & 700 & 25 &  20&  0.05 & 0.68 & 850 & $0.7034$ & $10^{-3}$\\
\hline
\hline
B20 & 0.7 & 2916 & 232 &  5&  0.05 & 0.69 & 120 & $0.7164$ & $10^{-3}$\\
\hline
B21 & 0.7 & 2916 & 232 &  3&  0.05 & 0.69 & 1650 & $0.7207$ & $4\times 10^{-4}$\\
\hline
C20 & 0.7 & 1372 & 109 &  10&  0.05 & 0.70 & 180 & $0.7165$ & $9.6\times 10^{-4}$\\
\hline
D20 & 0.7 & 704 & 56 &  20&  0.05 & 0.69 & 270 & $0.7165$ & $1.2\times
10^{-3}$\\
\hline
D21 & 0.7 & 704 & 56 &  20&  0.05 & 0.69 & 850 & $0.7185$ & $1.05\times
10^{-3}$\\
\hline
E20 & 0.7 & 352 & 28 &  40&  0.05 & 0.69 & 100 & $0.7143$ & $2\times
10^{-3}$\\
\hline
\hline
\end{tabular}
\caption{Différentes séries d'échantillons. $N$ est le nombre total de billes, $p$ la proportion en volume de
grosses billes, $N_2$ leur nombre, $M$ le nombre d'échantillons. L'étape LS donne lieu à  $n_{coll}$ collisions
par particule pour compacter de $\Phi_0$ à  $\Phi_1$, et la compacité finale 
obtenue, après l'étape de calcul par DM (à partir de $\Phi_1$)
 est $\langle \Phi^* \rangle$ en moyenne, l'écart-type étant $\Delta \Phi^*$.
\label{tab:liste}}
\normalsize
\end{table}
La figure~\ref{fig:coupe} représente la vue en coupe d'un échantillon
de 2916 billes (dont 232 grosses) dans un état de maximum de la
compacité, dont la figure~\ref{fig:paquet} donne une vue en perspective.
Les résultats de compacité sont présentés sur la figure~\ref{fig:phi}, en fonction de $N^{-1/2}$ (qui donne
des variations à peu près linéaires de la moyenne et de l'écart-type de $\Phi^*$ selon les 
relations~\ref{eqn:dphi} et~\ref{eqn:wphi}).
Les tailles d'échantillon étudiées sont tout aussi représentatives avec le mélange binaire que dans le cas
monodispersé~: on ne constate ni une dépendance en $N$, ni un niveau de fluctuations de $\Phi^*$ sensiblement plus
élevé dans le mélange. Suite à un réglage progressif des paramètres de l'étape L-S, le nombre de collisions
par particule lors de cette phase d'agitation et gonflement est variable entre les différentes séries
d'échantillons (tableau~\ref{tab:liste}). Toutefois, une tendance se dégage nettement~: plus l'étape d'agitation
se prolonge dans le processus d'assemblage, plus la compacité finale sera élevée -- résultat encore analogue au
cas monodispersé. 

Au-delà de la seule valeur de la compacité, la simulation numérique -- en particulier lorsque l'on poursuit les
calculs, comme nous l'avons fait ici, jusqu'à obtenir de bons états d'équilibre mécanique -- fournit bien 
d'autres informations. On a ainsi accès au réseau des contacts et aux forces qu'ils transmettent.
Le tableau~\ref{tab:cont1} donne les différents nombres de coordination $z_{ij}$, définis comme le nombre
moyen de contacts d'une particule d'espèce $i$ ($i=1$ pour les petites, $i=2$ pour les grosses) avec des
voisines d'espèce $j$\footnote{On a bien sûr 
$$
\frac{z_{21}}{z_{12}}=\frac{1-p}{p}\left(\frac{D}{d}\right)^3
$$}. Une certaine proportion $x_0$ des $N_1 = N-N_2$ petites billes ne transmet aucune force, alors que toutes les grosses sont sollicitées (c'est-à-dire que
les contraintes d'exclusion stérique sont actives).  
D'autre part, 
le tableau~\ref{tab:cont2} donne les valeurs en unités de $Pd^2$, $P$ étant la pression appliquée, des moyennes
des forces $F_{ij}$ et des rapports $F^{(2)}_{ij}$ 
entre écarts-types et moyennes (caractéristiques de la largeur de la
distribution, $\frac{\Delta F}{\langle F \rangle}$) pour les contacts
entre une bille d'espèce $i$ et une autre d'espèce $j$. 
\begin{table}[!hbt]
\centering
\begin{tabular}{|c||c|c|c|c|c|c|}  \cline{1-7}
Code 
& $z_{11}$ 
& $z_{12}$ 
& $z_{21}$ 
& $z_{22}$
& $z^*$
& $x_0$\\
\hline
\hline
A00 
& 
&  
&  
&  
& 6.074
&1.3\%\\
\hline
\hline
B10 
& $4.11\pm 0.03$ 
& $0.877\pm 0.003$
& $23.7\pm 0.8$ 
& $2.34 \pm 0.05$
& 6.068
&5.6\%\\
\hline
B11 
&$4.08\pm 0.04$ 
&$0.876\pm 0.014$ 
&$23.7\pm 0.1$ 
&$2.48\pm 0.1$
&6.064
&6.1\%\\
\hline
C10 
& $4.08\pm 0.06$ 
& $0.890\pm 0.018$
& $24.0\pm 0.5$ 
& $2.24\pm 0.19$
& 6.061
&5.7\%\\
\hline
C11 
&$4.12\pm 0.04$ 
& $0.885\pm 0.023$
& $23.9\pm 0.6$ 
&$2.47\pm 0.12$ 
& 6.089
&5.4\%\\
\hline
D10 
& $4.09\pm 0.05$ 
& $0.884\pm 0.026$
& $23.9\pm 0.$7 
&$2.30\pm 0.3$ 
& 6.058
&5.5\%\\
\hline
\hline
B20 
&$2.76\pm 0.05$ 
& $1.35\pm 0.015$
& $15.6\pm 0.2$ 
&$3.33\pm 0.05$ 
& 6.055
&13.6\%\\
\hline
B21 
& $2.65\pm 0.05$ 
& $1.33\pm 0.02$
& $15.3\pm 0.2$ 
&$3.73\pm 0.13$ 
& 6.053
&15.7\%\\
\hline
C20 
&$2.74\pm 0.04$ 
& $1.37\pm 0.02$
& $15.9\pm 0.2$ 
&$3.44\pm 0.10$ 
& 6.079
&13.6\%\\
\hline
D20 
& $2.62\pm 0.07$ 
& $1.38\pm 0.04$
& $15.9\pm 0.5$ 
&$3.48\pm 0.25$ 
& 6.045
&14.7\%\\
\hline
D21 
&$2.66\pm 0.11$ 
& $1.37\pm 0.04$
& $15.9\pm 0.4$ 
&$3.53\pm 0.23$ 
& 6.049
&14.4\%\\
\hline
E20 
&$2.70\pm 0.15$ 
& $1.36\pm 0.06$
& $16\pm 0.7$ 
&$3.37\pm 0.32$ 
& 6.043
&14.0\%\\
\hline
\hline
\end{tabular}
\caption{Nombres de coordinations, proportion $x_0$ de petites billes inactives. Notations définies dans le texte.
\label{tab:cont1}
}
\end{table}
Ces données concernant le réseau des contacts, conformément aux
résultats pour la compacité, indiquent que les assemblages granulaires
obtenus, hormis une légère influence de la durée de la
phase d'agitation, peuvent bien être considérés comme préparés dans le même état,
dont l'échantillonnage statistique est satisfaisant, indépendamment de
N dans l'intervalle étudié.
\begin{table}[!hbt]
\centering
\begin{tabular}{|c||c|c|c|c|c|c|}  \cline{1-7}
Code & $F_{11}$ &$F_{12}$ & $F_{22}$ & $F^{(2)}_{11}$ & $F^{(2)}_{12}$ & $F^{(2)}_{22}$\\
\hline
\hline
B10 & 0.827 &1.24 &4.47 &1.32 &1.36 &1.14  \\
\hline
B11 &0.780 &1.24 &4.88 &1.36&1.41&1.15 \\
\hline
C10 &0.823 &1.25 &4.51 &1.33 &1.38 &1.17  \\
\hline
C11 &0.796&1.21 &4.80 &1.34 &1.40 &1.15  \\
\hline
D10 &0.829 &1.24 &4.51 &1.33 &1.36 &1.15  \\
\hline
\hline
B20 &0.960 &1.43 &4.93 &1.45 &1.51 &1.18  \\
\hline
B21 & 0.800 &1.30 &5.70 &1.55 &1.62 &1.19  \\
\hline
C20 &0.951 &1.41 &4.81 &1.44 &1.49 &1.19  \\
\hline
D20 &0.931 &1.38 &5.08 &1.45 &1.52 &1.17  \\
\hline
D21 &0.901 &1.36 &5.18 &1.45 &1.52 &1.18  \\
\hline
E20 &0.953 &1.39 &5.04 &1.44 &1.49 &1.18  \\
\hline
\hline
\end{tabular}
\caption{Forces moyennes, écart-type réduit des distributions de force, par familles de
contacts. Notations définies dans le texte.
\label{tab:cont2}
}
\end{table}
Dans la limite des faibles pressions ou des billes rigides, on doit avoir~\cite{JNR2000} un nombre de
coordination moyen $z^*$, évalué avec les seules sphères qui portent des forces, égal à 6~:
\be
z^*=\frac{N_1}{N-x_0N_1}(z_{11}+z_{12})+\frac{N_2}{N-x_0N_1}(z_{21}+z_{22}) = 6,
\label{eqn:iso}
\ee 
correspondant à un réseau de contacts isostatique. Les résultats du tableau~\ref{tab:cont1} sont proches de
satisfaire cette condition, ce qui montre que la géométrie d'un assemblage de grains très faiblement déformables
en contact est décrite avec précision.

Le tableau~\ref{tab:cont1} montre aussi que les échantillons préparés
avec une plus longue étape d'agitation (ceux dont le code se termine
par <<~1~>>, voir le tableau~\ref{tab:liste}), et qui possèdent
une compacité plus élevée, présentent apparemment certaines prémices de ségrégation~: le nombre de coordination $z_{22}$ est
plus grand, signe que les grosses particules se rassemblent. Il conviendrait bien sûr, au-delà de cette étude préliminaire, d'étudier ce phénomène de plus près, 
et d'examiner si la tendance peut être poussée plus avant. Ces
mêmes échantillons se distinguent également par un rôle accru des
contacts entre gros grains dans la transmission des forces~: la valeur
de $F_{22}$ (tableau~\ref{tab:cont2}) montre que ces contacts sont non
seulement plus nombreux, mais aussi davantage sollicités.

La force normale de contact $f_{kl}$ entre le grain $k$ et son voisin $l$, dans la présente approche, sont des grandeurs
\emph{géométriques}, des multiplicateurs de Lagrange associés à la contrainte d'impénétrabilité.  Si le rayon
du grain $k$ subit une petite variation $\delta R$, alors le volume du système tout entier, si la liste des
contacts reste inchangée (ce qui est toujours possible dans un échantillon fini avec un $\delta R$ assez petit), 
varie au premier ordre de 
\be
\delta V =  \sum _l \frac{f_{kl}}{P} \delta R.
\label{eqn:varr}
\ee
De plus, le rapport $f_{kl}/P$ dépend dans la limite des grains rigides de la seule géométrie du paquet de billes en contact, 
du fait de l'isostaticité de cette structure~\cite{JNR2000}.
On conçoit donc que ce type de résultats puisse fournir des
indications pour la recherche de granulométries optimales,
l'équation~\eqref{eqn:varr} pouvant donner une idée de l'effet d'une
variation de la distribution des diamètres. En particulier, il est
bien sûr possible d'augmenter la taille des billes (au nombre de
$x_0N_1$) qui ne portent aucune force. 
De telles approches discrètes permettraient d'explorer le sens précis de diverses 
notions (<<~classe granulaire dominante~>>, <<~effet de paroi~>>, <<~effet de desserrement~>>...) introduites dans le
modèle de de Larrard~\cite{FdL99}, utilisé pour la formulation des ciments et des bétons.
\section{Conclusion et perspectives \label{sec:persp}}
Nous avons établi que la méthode de simulation utilisée permet d'obtenir de bons échantillons représentatifs pour les
mélanges binaires considérés. 
La détermination de la compacité est relativement aisée, même pour des échantillons de quelques centaines de grains,
lorsque le nombre de gros est de l'ordre de la dizaine. 
La compacité, pour le mélange binaire comme dans le cas monodispersé, augmente lorsque le processus
d'assemblage comporte une phase d'agitation prolongée. Cet effet, bien que relativement modéré, 
se dégage nettement des marges d'incertitude statistiques.
La compacité maximale usuelle correspond alors à la limite des assemblages rapides. 

L'étude des état de compacité maximale peut être poursuivie et généralisée à diverses granulométries, en particulier grâce aux performances croissantes
des ordinateurs et au calcul parallèle (pour lequel le LMSGC s'est doté d'une machine spécifique). Il sera particulièrement intéressant de relier certaines valeurs
caractéristiques de la compacité (valeur maximale, valeur critique...) à diverses propriétés rhéologiques (dilatance, viscosité, frottement interne),
éventuellement couplées à la ségrégation, des assemblages de grains seuls ou en présence d'un fluide. 
La simulation numérique des suspensions et des pâtes granulaires (suspensions très concentrées) est l'objet de recherches
en cours au LMSGC (A. Lemaître, A. Hammouti, J.-N. Roux, P.-E. Peyneau) et au LCPC-Nantes (N. Roquet). 

\renewcommand{\bibname}{R\'ef\'erences}

\end{document}